\documentclass[useAMS,usenatbib,twocolumn]{mn2e}
\usepackage[dvips]{graphicx}
\usepackage{amsmath}
\usepackage{amsfonts}
\usepackage{amssymb}

\newcommand{\multi}{{\sc MultiNest}}

\title[Eccentricity distribution as a Beta distribution]{Parametrizing the exoplanet eccentricity distribution with the Beta distribution}
\author[David M. Kipping]{David M. Kipping$^{1,2}$\thanks{E-mail:
dkipping@cfa.harvard.edu}\\
$^{1}$Harvard-Smithsonian Center for Astrophysics, 60 Garden St., Cambridge, MA 02138, USA \\
$^{2}$Carl Sagan Fellow}
\begin{document}

\date{Accepted 2013 June 3. Received 2013 May 22; in original form 2013 April 12}

\pagerange{\pageref{firstpage}--\pageref{lastpage}} \pubyear{2013}

\maketitle

\label{firstpage}

\begin{abstract}

It is suggested that the distribution of orbital eccentricities for extrasolar
planets is well-described by the Beta distribution. Several properties of the
Beta distribution make it a powerful tool for this purpose. For example, the
Beta distribution can reproduce a diverse range of probability density 
functions (PDFs) using just two shape parameters ($a$ and $b$). We argue that 
this makes it ideal for serving as a parametric model in Bayesian comparative 
population analysis. The Beta distribution is also uniquely defined over the
interval zero to unity, meaning that it can serve as a proper prior for 
eccentricity when analysing the observations of bound extrasolar planets. Using 
nested sampling, we find that the distribution of eccentricities 
for 396 exoplanets detected through radial velocity with high signal-to-noise 
is well-described by a Beta distribution with parameters 
$a = 0.867_{-0.044}^{+0.044}$ and $b = 3.03_{-0.16}^{+0.17}$. The Beta 
distribution is shown to be 3.7 times more likely to represent the underlying 
distribution of exoplanet eccentricities than the next best model: a Rayleigh + 
exponential distribution. The same data are also used in an example population 
comparison utilizing the Beta distribution, where we find that the short- and 
long-period planets are described by distinct Beta distributions at a confidence 
of 11.6\,$\sigma$ and display a signature consistent with the effects of tidal 
circularization.

\end{abstract}


\section{Introduction}
\label{sec:intro}

Thanks to the tireless efforts of observers in recent years, there now exists a 
sizeable library of orbital eccentricities ($e$) for extrasolar planets. 
Although photometric techniques are starting to emerge for measuring $e$, such
as Multibody Asterodensity Profiling (MAP) \citep{map:2012}, the precise
determination of this quantity has been historically determined by the radial
velocity variations (RV) of the host stars. 

This library of $e$ values has several uses and we focus on two particularly
useful applications here. The first is that the distribution can be exploited 
to test and refine theories of planet formation and evolution and offers a 
window into the possible scattering history of planetary systems 
\citep{rasio:1996,juric:2008,chatterjee:2008}. Such tests typically operate by 
taking a theoretical prediction for the distribution of various exoplanet 
parameters, in particular $e$, and comparing to the measured distribution from,
say, RV surveys. This comparison of distributions can also be 
extended to subpopulations of exoplanets, such as seeking evidence of tidal 
circularization by comparing the distribution of $e$ between short- and 
long-period planets. To make a quantitative comparison, one may use the
popular non-parametric and frequentist Kolmogorov-Smirnov (KS) test between the
two populations. Alternatively, a parametric approach (useful for Bayesian 
analyses) would be to regress one or more analytic distributions to the 
observed one. The parameters describing the analytic distribution may then be 
compared to test for statistically significant differences, or lack thereof.

A second useful application of an observed eccentricity distribution is that it
can be used to derive an informative prior on eccentricities in general. Before
the availability of this information, observers have been forced to adopt
uninformative priors, typically being a uniform prior over $0\leq e<1$, but an
informative prior can be preferable in many situations. Some examples we
consider are fitting RV data with phase gaps (which can lead to
spurious eccentricities), non-detection radial velocities used to place upper 
limits on $M_P\sin i$ (e.g. Kepler-22b; \citealt{borucki:2012}), blend analyses 
of transits requiring some eccentricity prior \citep{fressin:2011} and fitting 
transit light curves with an absence of any empirical eccentricity constraints. 
Using an informative prior naturally includes an observer's experience of the 
known distribution, taking into account whether a particular solution is a 
surprisingly rare answer or a very typical one. Any prior of course requires 
a parametrization of the observed eccentricity distribution. Furthermore, for 
use as a prior, the distribution should not reproduce negative eccentricities or
hyperbolic orbits (since any periodic transit, RV, asterometric, etc.
signal cannot result from such an orbit) and should integrate to unity over
the range $0\leq e<1$ to be defined as a proper prior.

From the aforementioned two major applications of the eccentricity distribution, 
we identify the following key requirements for any such parametrized probability
density function (PDF), $\mathrm{P}(e)$:

\begin{itemize}
\item[{\tiny$\blacksquare$}] $\mathrm{P}(e)$ should be defined over the range 
$0\leq e<1$ only i.e. no hyperbolic orbits or negative eccentricities
\item[{\tiny$\blacksquare$}] For a proper prior we require 
$\int_{e=0}^{1}\mathrm{P}(e)\,\mathrm{d}e=1$ i.e. the distribution is normalized 
over the defined range
\item[{\tiny$\blacksquare$}] We require $\mathrm{P}(e)$ to be able to reproduce 
a wide range of plausible distributions and be as efficient as possible i.e. use 
few parameters
\item[{\tiny$\blacksquare$}] The inverse of the cumulative density function 
(CDF) may be easily computed to serve as a practical (i.e. computationally
efficient) prior for direct sampling
\end{itemize}

\section{The Beta Distribution}
\label{sec:beta}

\subsection{Properties}

The Beta distribution, $\mathrm{P}_{\beta}(e;a,b)$, is a member of the 
exponential family defined over the range $0\leq e<1$ and satisfies all of 
the desired criteria described in the previous section. The functional form is
expressed in terms of either Gamma functions, or equivalently the Beta 
function, as

\begin{align}
\mathrm{P}_{\beta}(e;a,b) &= \frac{ \Gamma(a+b) }{ \Gamma(a) \Gamma(b) } e^{a-1} 
(1-e)^{b-1},\nonumber \\
\qquad&= \frac{ 1 }{ \mathrm{B}(a,b) } e^{a-1} (1-e)^{b-1}.
\end{align}

\begin{figure}
\begin{center}
\includegraphics[width=8.4 cm]{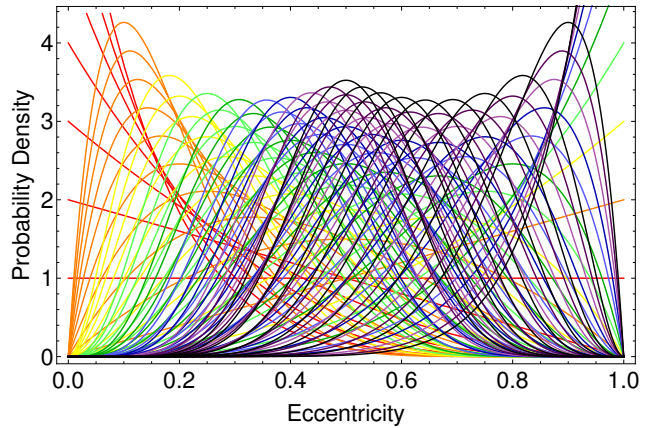}
\caption{\emph{Examples of the Beta probability density function, 
$\mathrm{P}_{\beta}(e;a,b)$, demonstrating the diverse range of distributions 
the function can produce. Going through red to purple and finally black, we 
explore from $a=1$ to $a=10$ in unity steps. For each colour we show 10 lines 
for $b=1$ to $b=10$ in unity steps.}} 
\label{fig:betaexamples}
\end{center}
\end{figure}

The first advantage of this form is that despite being described by just two 
parameters, $\mathrm{P}_{\beta}(e;a,b)$ is able to produce a wide and diverse 
range of probability distributions, as illustrated in 
Fig.~\ref{fig:betaexamples}. Secondly, the fact that the distribution is 
defined over the range zero to unity means it is suitable as a proper prior 
and it is trivial to show that

\begin{align}
\int_{e=0}^1 \mathrm{P}_{\beta}(e;a,b) = 1.
\end{align}

Thirdly, $\mathrm{P}_{\beta}(x;a,b)$ is clearly efficient given that only two 
parameters ($a$ and $b$) reproduce the wide range of distributions illustrated 
in Fig.~\ref{fig:betaexamples}. Finally, it may be shown that the CDF may be
inverted as a stable function, which is a requirement for using the Beta
distribution as a prior via direct sampling. The CDF is given by

\begin{align}
\mathrm{C}_{\beta}(e;a,b) &= \frac{\mathrm{B}(e;a,b)}{\mathrm{B}(a,b)} = I_e(a,b).
\end{align}

The inverse function is simply expressed $e=I_z^{-1}(a,b)$. A Beta distribution
prior can therefore be invoked by generating $z$ as a random uniform number 
between zero and unity and computing $e$, thus directly sampling from the prior
distribution. This inverse function is widely available in standard programming 
libraries. We note that \citet{hogg:2010} used a Beta distribution to model the
eccentricity distribution of a synthetic population but did not discuss how well
the distribution matches the observed distribution nor its potential as a prior.

\subsection{Comparison to other commonly used distributions}

One of the most commonly used PDFs for modelling the 
distribution of exoplanet eccentricities is a mixture between a Rayleigh
distribution and an exponential distribution (e.g. \citealt{steffen:2010,
wang:2011,map:2012}). The appeal of this mixture is that Rayleigh scattering 
reflects the effects of planet-planet scattering and the exponential component 
reflects the effects of tidal dissipation \citep{rasio:1996}. The associated
PDF is

\begin{align}
\mathrm{P}_{\mathrm{Rayleigh}}(e;\alpha,\lambda,\sigma) &= \alpha \lambda \exp\Big[-\lambda e\Big] + 
\frac{e (1-\alpha)}{\sigma^2} \exp\Big[-\frac{e^2}{2\sigma^2}\Big],
\label{eqn:rayleigh}
\end{align}

where $\alpha$ gives the relative contributions of the two PDFs, $\lambda$ is 
the width parameter of the exponential distribution and
$\sigma$ is the scale parameter of the Rayleigh distribution.

A major problem with the distribution of equation~\ref{eqn:rayleigh} is that
hyperbolic orbits ($e\geq1$) have a non-zero probability. This is true for both
the Rayleigh and exponential components taken individually too. Hyperbolic 
orbits (i.e. ejected planets) surely do naturally result from planet-planet 
scattering and planet-synthesis simulations may benefit from using this 
distribution \citep{rasio:1996}. However, it is not appropriate to use such 
a distribution as a prior for fitting, say, the RV time series 
of an exoplanet. This is because the very fact that a periodic planet signal
has been observed precludes $e\geq1$.

\citet{wang:2011} also used a uniform + exponential distribution to serve as
a null-hypothesis against the presence of a Rayleigh + exponential 
distribution. As before, for the purpose of serving as a prior in fitting, the 
exponential component will reproduce unobservable scenarios.

Another example of a model used recently for exoplanet eccentricities comes
from \citet{shen:2008} (hereafter ST08), who used a PDF requiring two shape 
parameters, $k$ and $a$.

\begin{align}
\mathrm{P}_{\mathrm{ST08}}(e;k,a) = \frac{1}{k} \Big(\frac{1}{(1+e)^a} - \frac{e}{2^a}\Big).
\label{eqn:shen}
\end{align}

It is easily shown that this distribution is not uniquely defined over the 
interval $0\leq e<1$.

\section{Example Regressions}
\label{sec:regressions}

\subsection{Regressing all planets}
\label{sub:fitall}

Regressing a PDF to a histogram of eccentricities is precarious in that the
results are sensitive to the chosen bin sizes. A more robust approach is to 
regress to the CDF which can be calculated at the smallest step sizes possible
i.e. the steps between each entry of the sorted list of eccentricities. As an 
example, we downloaded the eccentricites for all planets (413) discovered via 
RV from www.exoplanets.org \citep{wright:2011} on April $4^{\mathrm{th}}$ 2013. 
We make a cut in RV semi-amplitude of $(K/\sigma_K)>5$ in order to eliminate low 
signal-to-noise detections, leaving 396 exoplanets. 

These eccentricities represent the maximum likelihood estimates of $e$ for each 
planet. \citet{hogg:2010} argue that using the actual posteriors of $e$ for each
planet allows for a more accurate determination of the underlying distribution.
Unfortunately a large, homogenous and comprehensive database of such posteriors
is not available and would require a global reanalysis, which is outside the
scope of this short letter. Therefore, we proceed to use the maximum likelihood 
estimators of $e$ but acknowledge the possibility that this may be a biased 
indicator \citep{hogg:2010}. Despite this, we still argue that using the Beta
distribution with the fitted parameters presented in this section is a better
description of reality that other distributions suggested for reasons described
in \S\ref{sec:beta}.

The 396-length vector of eccentricities is first sorted from low to high.
Duplicate entries are removed to create a vector representing the minimum
step sizes in the CDF. For each entry in this vector, we then count the
number of entries in the original eccentricity vector which have a value
less than or equal to this. Normalizing by the total normal of entries
provides the probability and thus the CDF array. For this example, we 
elected the simple approach of computing errors for each array entry 
using Poisson counting statistics.

For the regression, we used the \multi\ package \citep{feroz:2008,feroz:2009},
which is a multimodal nested sampling algorithm \citep{skilling:2004}. 
\multi\ not only finds the maximum likelihood shape parameters and their
associated posterior distributions, but also computes the Bayesian evidence
of each model regressed. This latter functionality obviates the need for
using the frequently employed KS test, since Bayesian model selection can be 
easily performed using the evidences. A major benefit of using a Bayesian
approach is that we essentially penalise models for using unnecessary
complexity i.e. a built-in Occam's razor.

\begin{figure*}
\begin{center}
\includegraphics[width=18.0 cm]{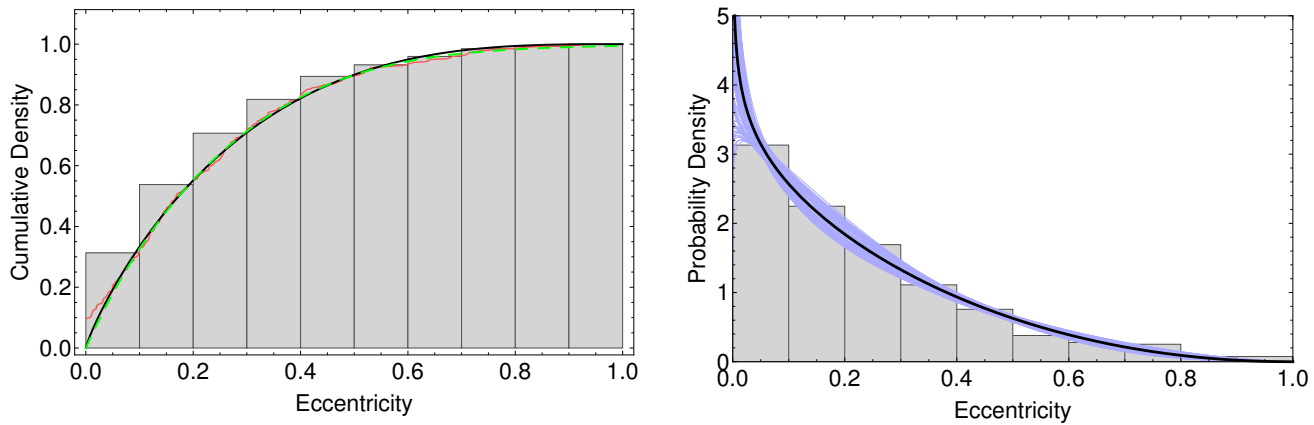}
\caption{\emph{\textbf{Left:} CDF of $e$ for 396 extrasolar planets
(grey bars), taken from www.exoplanets.org \citep{wright:2011}. The red line 
shows the smallest step size CDF. The black solid line shows the fitted Beta 
distribution and the green dashed line shows the fitted Rayleigh+exponential 
distribution. Although there is negligible difference between the latter two,
the Beta distribution requires one less shape parameter and is preferred
in a Bayesian sense. \textbf{Right:} PDF of the same
data (grey bars). We also show 100 random draws (blue lines) from the 
joint posterior of the Beta distribution parameters and the best fit in 
solid black. The range between the coloured lines illustrates the model 
uncertainty.}} 
\label{fig:cdffit}
\end{center}
\end{figure*}

For the parameter priors, we adopt modified Jeffrey's priors for $a$ and $b$ 
over the range $0$ to $10^2$ with an inflection point at unity to aid in quickly
scanning parameter space. After performing the regression, we derive $a = 
0.867_{-0.044}^{+0.044}$ and $b = 3.03_{-0.16}^{+0.17}$ (see 
Fig.~\ref{fig:cdffit}), where we quote median values and the 68.3\% credible 
intervals.

For comparison, other models were attempted starting with a simple uniform
distribution with two free parameters, $e_{\mathrm{min}}$ and 
$e_{\mathrm{max}}$. We directly sample from uniform priors in 
$e_{\mathrm{min}}$-$e_{\mathrm{max}}$ parameter space, except those cases 
where $e_{\mathrm{min}}>e_{\mathrm{max}}$.
Next, we regressed the popular Rayleigh + exponential distribution 
(equation~\ref{eqn:rayleigh}) using a modified Jeffrey's prior on 
$\lambda$ and $\sigma$ between $0$ and $10^2$ with an inflection point at 
unity. The prior for $\alpha$ was uniform over the interval zero to unity. 
We also tried a uniform + exponential, where we fixed $e_{\mathrm{min}} = 0$ 
and fitted $e_{\mathrm{max}}$ as a uniform prior over the interval zero to 
unity. $\alpha$ and $\sigma$ were treated as before.
Finally, we tried the intuitive model of ST08 provided in 
equation~\ref{eqn:shen}. For both $a$ and $k$, we used a modified Jeffrey's
prior between $0$ and $10^2$ with an inflection point at unity.

As the results show in Table~\ref{tab:fits}, the preferred model
we regressed to the data was that of a Beta distribution. The
Beta distribution is favoured over the next best model (the Rayleigh 
+ exponential distribution) with an odds ratio of 3.7 i.e. the Beta 
distribution is 3.7 times more likely to represent the underlying 
distribution. As already mentioned, the Beta distribution is defined 
over the interval zero to unity, unlike the other distributions 
attempted and is therefore favourable for use as a prior in subsequent 
analyses too.

\begin{figure}
\begin{center}
\includegraphics[width=8.4 cm]{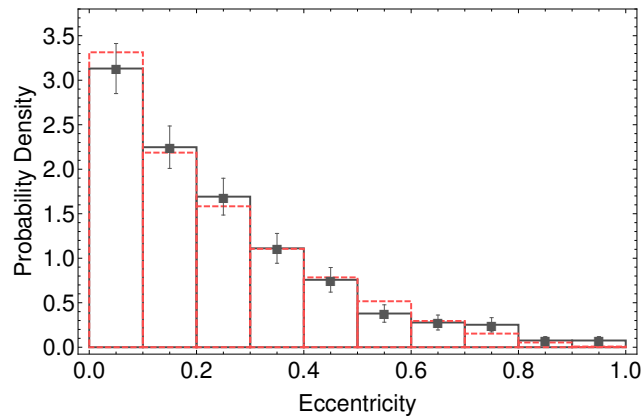}
\caption{\emph{Probability density distribution of $e$ for 396 extrasolar planets
(black bars), taken from www.exoplanets.org. The error bars shown are computed
using Poisson counting statistics. The red-dashed histogram shows a PDF of a 
synthetic population generated using the maximum likelihood parameters of a Beta
distribution regressed to the observed sample. Using just two shape parameters,
the fitted Beta distribution is fully consistent with the observed 
distribution.}}
\label{fig:pdffit}
\end{center}
\end{figure}

Using the maximum likelihood parameters of $a$ and $b$, we generated
a synthetic population of $10^5$ exoplanet eccentricities, which one would
hope to reproduce the observed distribution. Indeed, in Fig.~\ref{fig:pdffit}, 
this can be seen to be true, with each bin of the observed PDF falling within 
$\sim1$\,$\sigma$ of the synthetic one. The Beta distribution is therefore
certainly an excellent description of the observed exoplanet eccentricity 
distribution.

\begin{table*}
\caption{\emph{Using the observed eccentricities of 396 exoplanets from
www.exoplanets.org, we display the results of regressing several
CDFs. Along with the parameters (columns 3-5) we also
show the Bayesian evidence (column 2) of each regression (higher is better).}} 
\centering 
\begin{tabular}{l l l l l} 
\hline\hline 
\textbf{Distribution} & \textbf{Evidence} & \textbf{Parameter 1} & \textbf{Parameter 2} & \textbf{Parameter 3} \\ [0.5ex] 
\hline
Uniform[$e_{\mathrm{min}}$,$e_{\mathrm{max}}$] & $-664.761\pm0.053$ & 
$0.90_{-0.66}^{+1.48}\times10^{-4}$ & $0.6071_{-0.0037}^{+0.0037}$ & - \\ %
Beta[$a$,$b$] & $+374.705\pm0.046$ & $0.867_{-0.044}^{+0.044}$ & 
$3.03_{-0.16}^{+0.17}$ & - \\
Rayleigh+Exp[$\alpha$,$\sigma$,$\lambda$] & $+373.400\pm0.049$ & 
$0.781_{-0.132}^{+0.083}$ & $0.272_{-0.036}^{+0.021}$ 
& $5.12_{-0.61}^{+1.44}$ \\
Uniform+Exp[$\alpha$,$\sigma$,$e_{\mathrm{max}}$] & $+332.506\pm0.054$ & $0.1292_{-0.070}^{+0.069}$ & $0.2229_{-0.0048}^{+0.0051}$ & $0.559_{-0.035}^{+0.037}$ \\
ST08[$k$,$a$] & $+371.475\pm0.051$ & $0.2431_{-0.0059}^{+0.0060}$ & $4.33_{-0.18}^{+0.18}$ & -  \\ [1ex]
\hline\hline 
\end{tabular}
\label{tab:fits} 
\end{table*}

\subsection{Population comparison example}
\label{sub:fitlocal}

Here, we show how population comparison may be achieved in a Bayesian sense
without the use of the frequentist KS test and easily modelled with the
Beta distribution. In this example, we consider two possible hypotheses which 
describe the underlying distribution of the eccentricity of exoplanet 
eccentricities:

\begin{itemize}
\item[{\tiny$\blacksquare$}] $\mathcal{H}_1$: The eccentricity of all 
exoplanets is described by a single Beta distribution,
$\mathrm{P}_{\beta}(a_{\mathrm{global}},b_{\mathrm{global}};e)$
\item[{\tiny$\blacksquare$}] $\mathcal{H}_2$: The eccentricity of the 
short-period exoplanets is described by a Beta distribution,
$\mathrm{P}_{\beta}(a_{\mathrm{short}},b_{\mathrm{short}};e)$, and
that of long-period planets by 
$\mathrm{P}_{\beta}(a_{\mathrm{long}},b_{\mathrm{long}};e)$
\end{itemize}

We define ``short period'' and ``long period'' planets by computing
the median period of the 396 exoplanets analysed in the previous
subsection. Two separate CDFs are generated, split by this median
period (382.3\,days). The CDFs are computed using the same method
described in \S\ref{sub:fitall}. The CDFs are then fitted with 
global shape parameters for hypothesis $\mathcal{H}_1$ and local shape 
parameters for hypothesis $\mathcal{H}_2$.

The results of this exercise are shown in Table~\ref{tab:betadual}. We
note that the global fit retrieves slightly different parameters than those 
found when using a single CDF function. Parameter $a$ is found to differ 
by 2.4\,$\sigma$ and $b$ by 1.8\,$\sigma$. We attribute this difference 
to the binning procedure where the number of unique eccentricities defines 
the maximum resolution possible when constructing a CDF. As a result, the
combined CDF result will have the higher resolution and thus greater
reliability.

\begin{figure*}
\begin{center}
\includegraphics[width=16.8 cm]{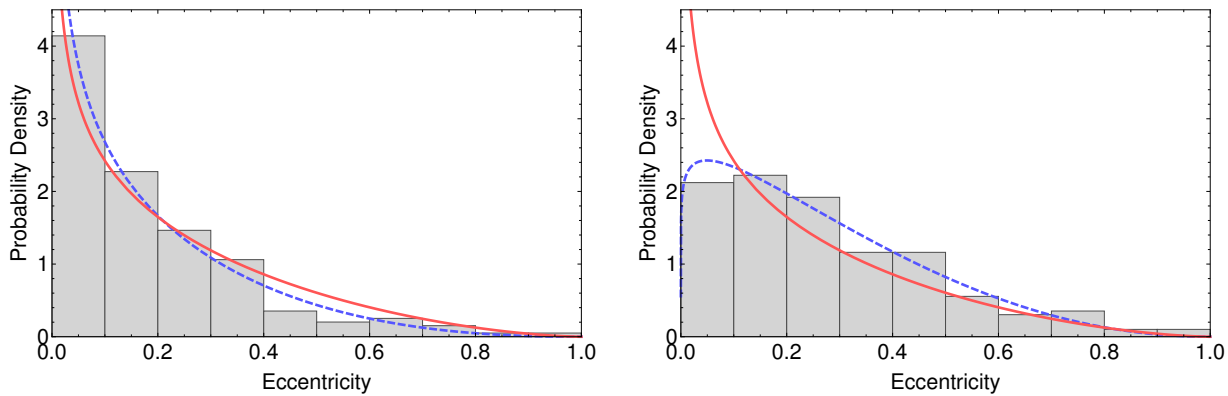}
\caption{\emph{Probability density function for short-period planets (left)
and long-period planets (right) drawn from the www.exoplanets.org archive.
The red solid line shows the result of regressing a single Beta distribution to
both sets. The blue dashed line is the result of regressing two independent 
Beta distributions.}} 
\label{fig:betadual}
\end{center}
\end{figure*}

The Bayesian evidence yields an 11.6\,$\sigma$ preference for hypothesis
$\mathcal{H}_2$. We therefore conclude that there is a significant
difference between the eccentricity distributions of short- and long-period
exoplanets. Furthermore, the short-period planets show a larger fraction of
low-eccentricity planets relative to the flatter distribution found for
long-period planets (see Fig.~\ref{fig:betadual}). This is consistent 
with the effects of tidal circularization \citep{rasio:1996}.

\begin{table*}
\caption{\emph{Comparison of regressing a global Beta distribution 
($\mathcal{H}_1$) versus two independent Beta distributions 
($\mathcal{H}_2$) to the short- and long-period exoplanets from
www.exoplanets.org. $\mathcal{H}_2$ is favoured at 11.6\,$\sigma$.}} 
\centering 
\begin{tabular}{l l l l l l l} 
\hline\hline 
\textbf{Hypothesis} & \textbf{Distribution} & \textbf{Evidence} & \textbf{Parameter 1} & \textbf{Parameter 2} & \textbf{Parameter 3} & \textbf{Parameter 4} \\ [0.5ex] 
\hline
$\mathcal{H}_1$ & Beta[$a_{\mathrm{global}}$,$b_{\mathrm{global}}$] & $264.528 \pm 0.044$ & $0.711_{-0.044}^{+0.049}$ & $2.57_{-0.17}^{+0.19}$ & - & - \\ %
$\mathcal{H}_2$ & Beta'[$a_{\mathrm{long}}$,$b_{\mathrm{long}}$,$a_{\mathrm{short}}$,$b_{\mathrm{short}}$] & $334.654 \pm 0.060$ & $1.12_{-0.10}^{+0.11}$ & $3.09_{-0.29}^{+0.32}$ & $0.697_{-0.481}^{+0.066}$ & $3.27_{-0.32}^{+0.35}$ \\ [1ex]
\hline\hline 
\end{tabular}
\label{tab:betadual} 
\end{table*}

\section{Discussion \& Conclusions}
\label{sec:discussion}

We have shown how the Beta distribution is a useful tool for parametrizing the
distribution of exoplanet orbital eccentricities. The Beta distribution 
is well suited for this purpose, thanks to its diverse range of PDFs using just
two shape parameters ($a$ and $b$), a strictly defined interval between $0$ and 
$1$ as expected for bound exoplanets, and possessing an easily invertible CDF
for the purpose of sampling from a Beta distribution prior.

By regressing the known CDF of orbital eccentricities from exoplanets
detected through the RV technique at www.exoplanets.org \citep{wright:2011}, we
have shown how the Beta distribution is 3.7 times more likely to represent
the underlying distribution of orbital eccentricites than the next best
competing model: that of a Rayleigh + exponential distribution (see
Table~\ref{tab:fits}). We find that the parameters 
$a = 0.867_{-0.044}^{+0.044}$ and $b = 3.03_{-0.16}^{+0.17}$
provide an excellent match to the data and are able to reproduce the
observed distribution (see Fig.~\ref{fig:pdffit}). We suggest that observers
may use these shape parameters to define an informative eccentricity prior.
Sampling from this prior will not only naturally include an observer's previous 
experience, but is also more computationally efficient since the distribution
is skewed to lower eccentricities where Kepler's transcendental equation is more 
expediently evaluated.

Finally, we have shown how the Beta distribution may be used for comparing
populations of exoplanet eccentricities, with an example application to
comparing short- and long-period planets. Here, we find that a two-population
model is strongly favoured at more than $11$\,$\sigma$ and we find that
short-period planets have a higher proportion of low-eccentricity planets
where long-period planets exhibit a flatter distribution, consistent with
tidal circularization (see Fig.~\ref{fig:betadual}).

\section*{Acknowledgements}

DMK has been supported by the NASA Carl Sagan Fellowships. Thanks to
Joel Hartman and Kevin Schlaufman for useful discussions in preparing this
manuscript. This research has made use of the Exoplanet Orbit Database and the 
Exoplanet Data Explorer at exoplanets.org.




\label{lastpage}

\end{document}